# Ultrafast dynamics in normal and CDW phase of 2H-NbSe$_2$

A. Anikin[1], R.D. Schaller[2], G.P. Wiederrecht[2], E.R. Margine[3], I.I. Mazin[4], G. Karapetrov[1,5]

[1] Department of Physics, Drexel University, Philadelphia, PA 19104
[2] Center for Nanoscale Materials, Argonne National Laboratory, Lemont, Illinois 60439
[3] Department of Physics, Applied Physics and Astronomy, Binghamton University-SUNY, Binghamton, New York 13902
[4] Department of Physics and Astronomy, George Mason University, Fairfax, VA22030
[5] Department of Materials Science and Engineering, Drexel University, Philadelphia, PA 19104

We investigate carrier and collective mode dynamics in 2H-NbSe$_2$ using time-resolved optical pump-probe spectroscopy and compare the results with first-principle calculations. Broadband ultrafast reflectivity studies of 2H-NbSe$_2$ in a wide temperature interval covering the normal, charge density wave (CDW) and superconducting phase were performed. Spectral features observed in the transient reflectivity experiment were associated with specific optical transitions obtained from band structure calculations. Displacive excitation of coherent phonons showed CDW-associated coherent oscillations of the soft phonon mode across the whole spectral range. Temperature evolution of this coherent phonon mode in the low-excitation linear regime shows softening of the mode down to the CDW transition temperature $T_{CDW}$ with subsequent hardening below $T_{CDW}$. The global fit of the broadband probe data reveals four different relaxation times associated with characteristic electron-electron, electron-phonon and phonon-phonon relaxation processes. From first principle calculations of electron-phonon coupling we associate the few picosecond electron-phonon relaxation time $\tau_2$ with a specific group of phonons with frequencies around 20 meV. On the other hand, the anomalously long relaxation time of $\tau_3 \sim 100$ ps is associated with anharmonicity-driven phonon-phonon scattering. All relaxation processes result from anomalies near the second order CDW phase transition that are reflected in the temperature dependencies of the characteristic relaxation times and amplitudes of optical densities. At highest fluences we observe electronic melting of the CDW and disappearance of the mode hardening below $T_{CDW}$.

## 1. Introduction

Layered transition metal dichalcogenides (TMD) are a large family of quasi-2D materials that often exhibit charge density waves (CDW) and/or superconductivity (SC). Both CDW and SC are usually a product of electron phonon coupling, but in several dichalcogenide systems more exotic interactions have been discussed. For example, TiSe$_2$ is still discussed as a potential excitonic insulator, although the jury is still out [1]. Lack of inversion symmetry in some TMDs leads to interesting effects such as potentially chiral CDW [2–4] and spin-momentum locking in single-layer NbSe$_2$ leading to Ising superconductivity. [5] Although the dichalcogenides have been extensively studied for over 50 years, many interesting topics remain open. Novel experimental methods that focus on high spatial and temporal resolution provide further insights into the interplay of electrons and phonons that leads to the CDW and SC phases. Femtosecond optical spectroscopy provides a unique opportunity to examine collective modes with superior frequency resolution and high dynamic range. [6] In this paper we explore femtosecond optical reflectivity and coherent phonon dynamics in 2H-NbSe$_2$ – material isostructural and similar to 2H-NbS$_2$ in that exhibits strong anharmonicity at low temperatures, but nevertheless goes through both CDW and superconducting phase transitions with both phases coexisting below 7.2 K. [7,8]

2H-NbSe$_2$ is a van der Waals material belonging to *P6$_3$/mmc* space group with trigonal Nb layer stacked between two trigonal Se layers having periodicity of double Se-Nb-Se trilayer. It remains metallic from room temperature down to its superconducting transition at $T_c = 7.2\ K$ [9] with three hole and two electron Fermi surfaces [10], with different electron phonon coupling, leading to a two-gap superconductivity [11] and to switching of the Hall conductivity from p-type at $T_H > \sim 18$ K to n-type below $T_H$. [12] At temperatures below 33.5 K an incommensurate CDW sets in, corresponding to simultaneous condensation of three CDWs with the wave vectors $q\{1,0,0\}$, $q\{0,1,0\}$, $q\{1,1,0\}$, where $q \approx 1/3$, close to a reconstruction with the 3x3 periodicity. [13] CDW-like fluctuations persist up to 100 K in bulk NbSe$_2$, well above the thermodynamic second order CDW phase transition temperature of 33.5 K. [14,15] Under pressure the CDW instability is rapidly suppressed, with only a minor effect on superconductivity, indicating that the latter is not due to the soft modes related to the CDW instability [16–19].

Although initially many studies linked the CDW formation to a simple Peierls instability and Fermi surface nesting, detailed band structure calculations [10,20,21] and experiments [22–24] show that electron-phonon coupling plays the determining role in condensing the CDW ground state [21,24]. Moreover, experimentally and theoretically determined Lindhart susceptibility function shows no peaks [24,25] and only a very broad maximum at the corresponding wave vector $q_{CDW}$, extending to almost half the Brillouin zone in the ΓM direction [22], which indicates a minor role played by electronic nesting. The amplitude of CDW charge supermodulation is close to 9%, resulting in a very small kink ( ~5% ) in the metallic temperature dependence of resistivity at $T_{CDW}$. [9,26,27] The theory of strong q-dependent electron-phonon coupling (EPC) was proposed as an attempt to explain the structural and electronic transition [20,28]. The calculations are consistent with the experimental phonon dispersion [25]. Experimentally, q-dependent strong EPC was confirmed by Weber et al. [22] and further studied in detail later on. [29–31]

An incommensurate CDW in NbSe$_2$ forms along the ΓM direction in reciprocal space with $q_{CDW} = \frac{2\pi}{3a}(1-\delta)$ with $\delta \approx 0.02$. The CDW gaps the inner Fermi surface cylinder around the K point of the first Brillouin zone, as first proposed theoretically [10,16,28,32] and shown experimentally [11,22,23,29]. The longitudinal optical phonon Σ branch couples to the Nb 4d electrons near K and K' pockets of the double-walled Fermi surface sheets, and, as we will discuss later in the paper, is instrumental for superconductivity, but not for the CDW. The charge ordering induces a negative carrier plasmon dispersion leading to a minimum at $q=q_{cdw}$ of the ~1eV plasmon, which represents an electronic signature of fluctuating CDW order parameter. [33,34]

Electron-phonon coupling cannot always stabilize CDW order even in materials that are structurally and electronically almost equivalent, such as NbSe$_2$ and NbS$_2$. It seems that the strength of the electron-phonon interaction has to overcome the anharmonic effects, as the temperature is lowered. In NbS$_2$, anharmonicity prevents condensation of CDW phase down to T=0 K [35], although the fluctuating CDW order parameter is present at low temperatures. [36] Anharmonic effects can be reduced by making the system more two-dimensional. Reduction of dimensionality, i.e. electronic decoupling between the van der Waals layers, reduces conductivity and charge screening [18,37], as well as prevents ionic charge transfer [38] between the transition metal cation and dichalogenide anions. Thus in isolated monolayer NbS$_2$ the CDW phase is stable and wins over the anharmonic effects below 5K. [39,40] NbSe$_2$ in a monolayer form experiences enhancement of $T_{CDW}$ to 145 K, which exceeds the onset temperature of the CDW fluctuations in the bulk system (~100 K). [41,42] Even in bulk NbSe$_2$, the surface layer seems to experience CDW fluctuations as high as room temperature, as observed by surface sensitive techniques like small angle inelastic x-ray scattering [43] or ARPES. [11] On the other hand, effects that enhance anharmonicity and reduce the anisotropy, such as hydrostatic pressure, suppress the $T_{CDW}$. [44,31]

In this paper we explore anharmonic processes in NbSe$_2$ using transient optical reflectivity. We explore the temperature evolution of the anharmonicity in order to understand the electronic processes taking place near the normal to incommensurate CDW second order phase transition in 2H-NbSe$_2$. We find that the femtosecond optical pulse leads to displacive excitation of a coherent phonon (DECP) with A$_1$ symmetry [45] that corresponds to the two-phonon peak observed in Raman spectroscopy [43,46–50]. We observe strong damping of the coherent phonon and correlate it with the broad spectral width of the Raman two-phonon peak. As temperature is lowered towards $T_{CDW}$, the phonon softens and the damping enhances, in agreement with expected strong electron-phonon coupling near the CDW transition in this system. Below the $T_{CDW}$ the coherent phonon hardens and dephasing time increases, effects that have not been previously observed with optical probes. We believe that blue-shifting of the coherent phonon below $T_{CDW}$ is a result of strengthening of the CDW order parameter as temperature is lowered away from $T_{CDW}$. The contribution of the phase modes, which are normally the acoustic modes of the incommensurate CDW state, is raised to a finite frequency if the inversion symmetry is broken by the layer stacking [51], as in NbSe$_2$. As the fluence of the pump pulse increases, electronic melting of CDW takes place and the blue-shift of the DECP below $T_{CDW}$ disappears thus confirming that the fragile CDW phase mode is involved in the process.

## 2. Experimental and computational details

High quality single crystals of 2H-NbSe$_2$ were grown by the chemical vapor transport method with iodine as the transport agent. Resistivity measurements revealed R(300K)/R(8K) ≈ 100, T$_C$ = 7.2 K and T$_{CDW}$ = 33.5 K, values similar to the highest quality crystals (see Supplementary Information). [52–55] A freshly cleaved surface of the single crystal was exposed before introducing the sample into the optical cryostat. Pump-probe reflectivity measurements were conducted using a linearly polarized pump laser pump at 800 nm (1.55 eV) operating at a repetition rate of 1 kHz and a pulse width of ~ 25 fs. The measured beam diameter is approximately 600 μm. Calculated pulse penetration depth is about 30 nm. To probe the dynamics, a linearly polarized white probe beam was used. The measurements were taken at temperature intervals from 3 K to 150 K at several pump fluences. Data analysis of the changes in optical density were extracted using the open-source software package Glotaran [56].

First-principles calculations were performed with density functional theory (DFT) using the Quantum ESPRESSO (QE) [57] code. We employed optimized norm-conserving Vanderbilt (ONCV) pseudopotentials [58,59] with the Perdew-Burke-Ernzerhof (PBE) exchange-correlation functional in the generalized gradient approximation [60], where the Nb $4s^24p^64d^35s^2$ and Se $4s^24p^4$ orbitals were included as valence electrons. All calculations were performed for the experimental lattice parameters at ambient pressure [22] with relaxed internal coordinates. We used a plane wave kinetic-energy cutoff value of 80 Ry, and the electronic and vibrational Brillouin zones were sampled using 24 × 24 × 8 and 6 × 6 × 2 points, respectively. A Methfessel and Paxton smearing [61] width of 0.025 Ry was used in order to resolve the CDW instability. The electron-phonon calculations were performed with the EPW code [62,63] of the QE distribution, in conjunction with the Wannier90 library. [64,65] We used 22 maximally localized Wannier functions (five d-orbitals for each Nb atom and three p-orbitals for each Se atom) and a uniform Γ-centered 12 × 12 × 4 electron-momentum grid. The linewidth associated with the electron-phonon interaction $\gamma_{\nu,\mathbf{q}}$ of mode ν with momentum **q** were evaluated on a uniform 240 × 240 × 60 **k**-point grid. The isotropic Eliashberg spectral function and the phonon linewidth spectral function were evaluated on uniform 80 × 80 × 40 **k**- and 40 × 40 × 20 **q**-point grids. The Dirac deltas were replaced by Lorentzians of width 25 meV (electrons) and 0.05 meV (phonons).

### 3. Results and discussion

Figure 1 displays the transient change in reflectivity, *R*, induced by the pump pulse, defined as signal $A(\omega, T) = -ln\left(R_{pump}(\omega, T)/R_{no\ pump}(\omega, T)\right)$. The data were collected at 3 K with 0.5 mJ/cm$^2$ pump fluence. The positive values on the color map correspond to a decrease in reflectivity (increased optical density) due to the pump pulse, while the negative values correspond to an increase in reflectivity (decreased optical density) induced by the pump pulse.

The physics of the transient response is, in the first approximation, rather simple. First, the laser pump excites electrons, by triggering electronic transitions with the energy equal to the laser photon energy. On a rather short time scale, $\tau_1$~0.3 ps, these electrons thermalize among themselves *via* electron-electron scattering. Typically, the resulting electronic temperature is $T_e$~1000 K. This, first of all, leads to a modification of the electronic absorption $\varepsilon_2(\omega) = const \cdot \sum_{\mathbf{k},i,j}(f_{k,i} - f_{k,j})\delta(E_{k,j} - E_{k,i} - \omega)$, where *f(T)* are Fermi occupation factors, and we have absorbed the average dipole matrix elements into the prefactor. Taking the derivative with respect to *T*, we immediately see that summation in now over the product of the two δ-functions, $\sum_{\mathbf{k},i,j}\delta(E_{k,j} - \mu)\delta(E_{k,j} - \omega)$, where μ is the Fermi energy. This is equivalent to integrating over the line where the two isoenergetic surfaces intersect (Fig. 2), with the weight equal to the inverse cross product of the relevant electron velocities. [66] The sign does not depend on whether the transition is to or from the Fermi level, but on the details of the joint density of states and matrix elements.

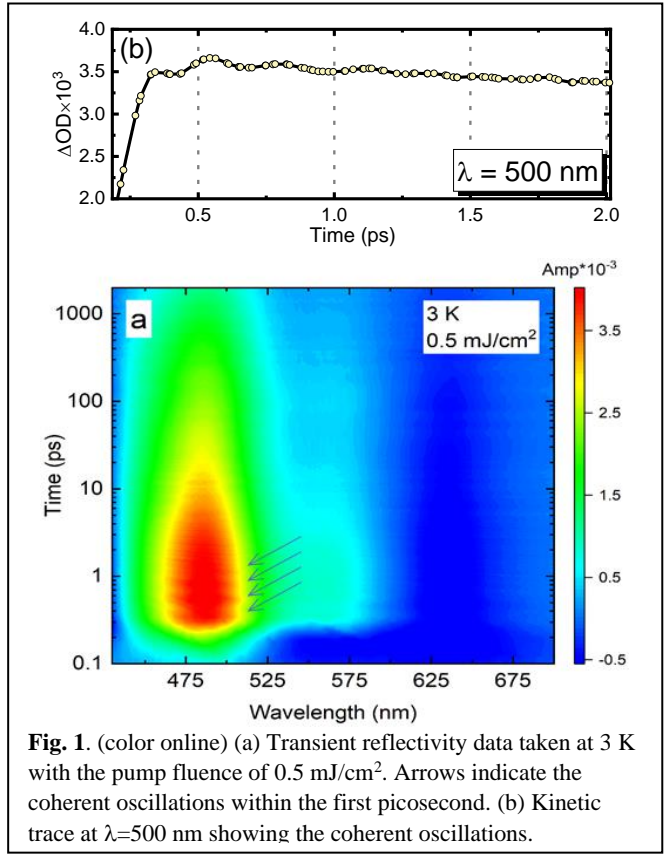

**Fig. 1**. (color online) (a) Transient reflectivity data taken at 3 K with the pump fluence of 0.5 mJ/cm$^2$. Arrows indicate the coherent oscillations within the first picosecond. (b) Kinetic trace at λ=500 nm showing the coherent oscillations.

Three features can be seen in Fig. 1. The strongest is at 485 nm (2.6 eV) and accompanied with a weaker one at 570 nm (2.18 eV) and an even weaker one, of the opposite sign, at 635 nm (1.95 eV). These wavelengths correspond to features that have been observed in static reflectivity. [67–69] The calculated reflectivity together with the optical absorption $\varepsilon_2(\omega)$ are shown in Fig. 3. The energy is scaled by 4% to assure better agreement with the experiment. [66–68] This is not uncommon in optical calculation, reflecting renormalization of the bandwidth due to strong interaction. For instance, in Ref. [70] the best agreement of the calculated optical absoprion was found when all features were red-sfifted by ~10%. This was ascribed to the band width renormalization due to magnon exchange. The effect is, apparently, considerably weaker in NbSe$_2$, consistent with the presence of spin fluctuation there [71], albeit, of course, much weaker than in CrO$_2$. We can clearly see a very strong and sharp peak exactly at the frequency of our strongest feature at 2.6 eV and a much weaker one at 2.18 eV. We also see a tiny shoulder corresponding to our third feature.

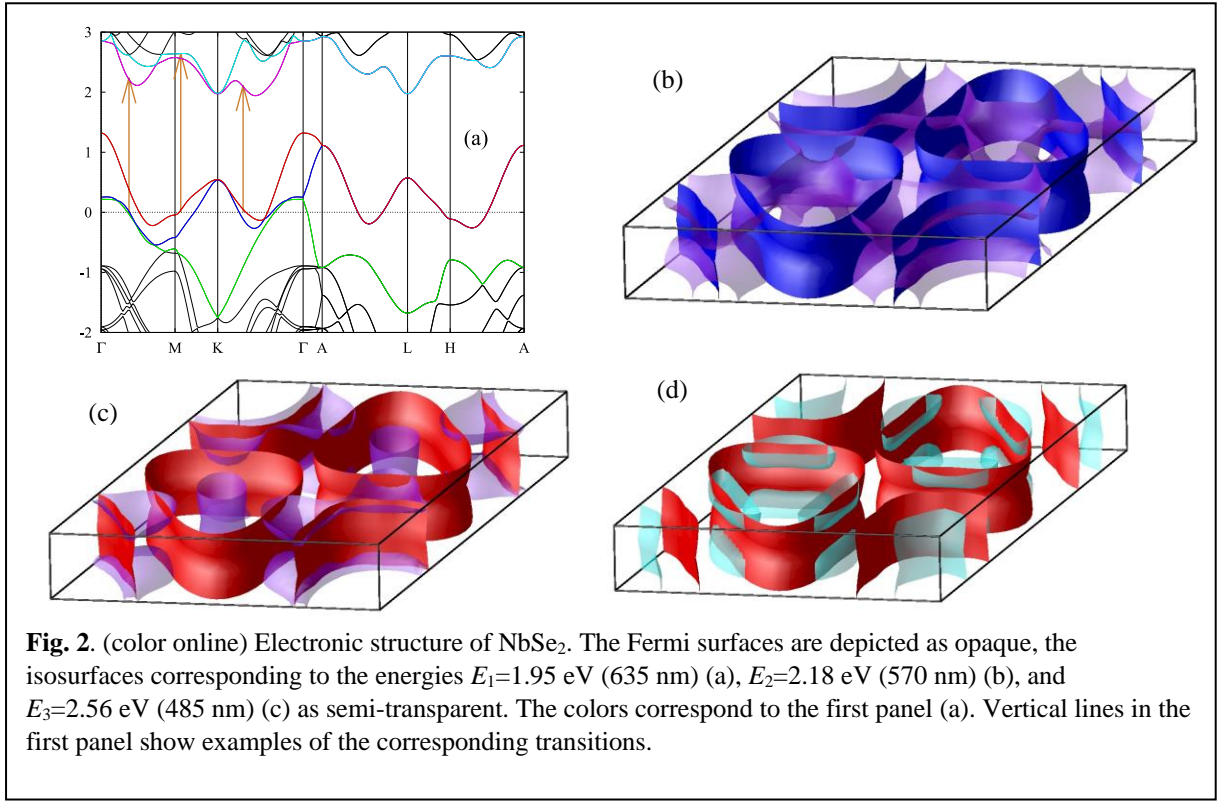

**Fig. 2.** (color online) Electronic structure of NbSe$_2$. The Fermi surfaces are depicted as opaque, the isosurfaces corresponding to the energies $E_1$=1.95 eV (635 nm) (a), $E_2$=2.18 eV (570 nm) (b), and $E_3$=2.56 eV (485 nm) (c) as semi-transparent. The colors correspond to the first panel (a). Vertical lines in the first panel show examples of the corresponding transitions.

Band analysis reveals that while the feature appears as combinations of several interband transitions, each one is dominated by the one transition, depicted in Fig. 2.

Another interesting piece of information that can be extracted from our data is the oscillatory behavior of the transient response. This is typical for displacive excitation of coherent phonons (DECP), whereupon nearly-instantaneous heating of the electron subsystem to $T_e$~1000 K shifts the equilibrium positions of the ions that are not fixed by symmetry (in our case, the $z$ position of Se), and then the ions oscillate around the new position with a pattern consistent with the $A_{1g}$ symmetry. We see no signature of the actual $A_{1g}$ phonon, which exists at ~230 cm$^{-1}$, presumably, because this phonon does not modify the dielectric response at 400-700 nm. However, we see a broad maximum centered at about 3.5-4.5 THz (120-150 cm$^{-1}$), exactly matching the well-known broad feature in the $A_{1g}$ Raman spectra [43,46–49,72], usually interpreted as a composite excitation of two phonons propagating in the opposite direction (a standing wave) [73,74]. This will be generated, for instance, if there is a lateral gradient of $T_e$. Generally, such a feature would be weak compared to a single $A_{1g}$ phonon, but if such a wave modulates the dielectric function in the relevant energy range, which is highly possible (because, as opposed to the $A_{1g}$ mode, it involved Nb displacements), it will be the only one observed.

Let us analyze this feature in more details (Fig.4). Highly damped coherent oscillations can be resolved within the first picosecond across the whole measured spectral interval. Oscillations were detected at all sample temperatures which allows us to study their temperature evolution. Since the oscillation amplitudes are relatively weak, we averaged the time traces from 500 nm to 560 nm. The traces were fitted by exponentially-modified peak function convoluted with damped sine function to capture the oscillations (Fig. 4a and 4b). Extracted temperature dependence of the frequency and dephasing rate are shown in Fig.4(c,d).

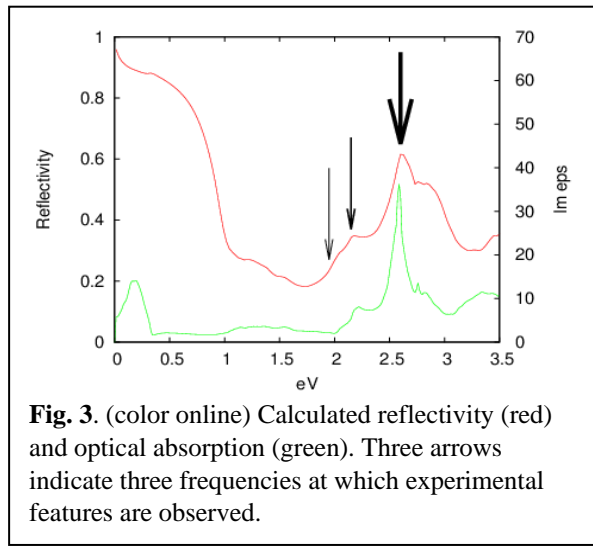

**Fig. 3.** (color online) Calculated reflectivity (red) and optical absorption (green). Three arrows indicate three frequencies at which experimental features are observed.

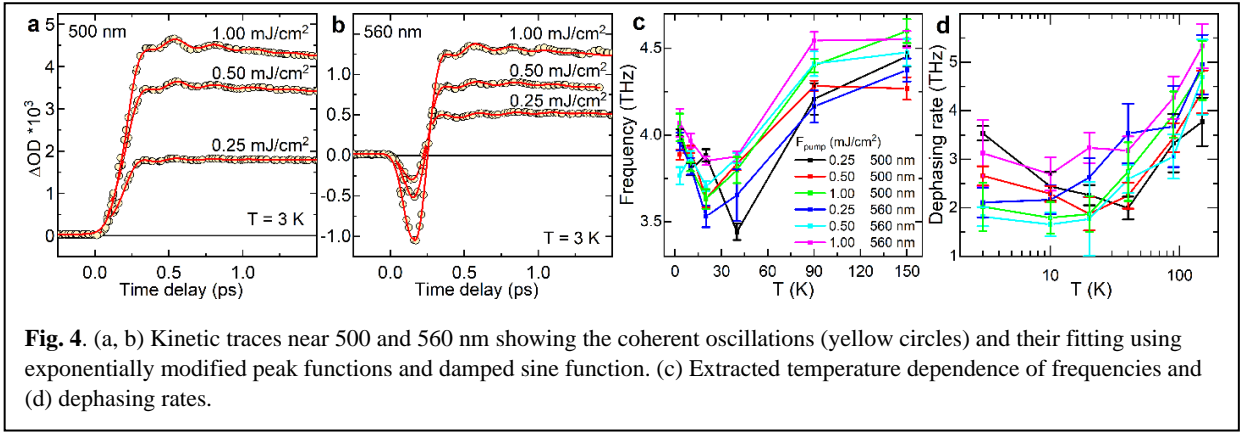

**Fig. 4**. (a, b) Kinetic traces near 500 and 560 nm showing the coherent oscillations (yellow circles) and their fitting using exponentially modified peak functions and damped sine function. (c) Extracted temperature dependence of frequencies and (d) dephasing rates.

We can clearly see softening of this composite excitation from 150 K down to $T_{CDW}$. At the same time, the dephasing rate gets slightly shorter. Below $T_{CDW}$ this excitation hardens and the damping rapidly decreases. However, the frequency remains finite at T=$T_{CDW}$. One can understand that by looking at the calculated phonon dispersion near $q = q_{CDW}$. Our calculations (Fig. 5), consistent with previously published results by Weber et al. [23], indicate that the soft mode frequency is minimal at $q_{CDW}$ and then increases quadratically as one moves away from $q_{CDW}$, and reaches a maximum of about 3 THz (12 meV), which is temperature-independent. The minimum at $q_{CDW}$, as measured by inelastic x-ray scattering [75], shifts with temperature in a mean-field manner as $\sqrt{\frac{T}{T_{CDW}} - 1}$, and raises about twice that fast again below $T_{CDW}$.

Approximating the frequency of this branch as

$$\omega(T, q) = 2A \frac{|q - q_{CDW}|^2}{q_{CDW}^2} + B \sqrt{\frac{T}{T_{CDW}} - 1} , \qquad (1)$$

where A~3 THz and B, from Ref. [75], ~0.8 THz, and averaging over $|q - q_{CDW}| < q_{max}$, we get $\omega(T, q) = 0.6A + B\sqrt{\frac{T}{T_{CDW}} - 1} \approx 2$ THz+$0.8\sqrt{\frac{T}{T_{CDW}} - 1}$ THz. Below $T_{CDW}$, B as extracted from Ref. [22], is about twice larger. As our results show (Fig.6), this scenario describes our data rather well, strongly suggesting that the two-phonon mode is composed of the phonons of the same branch that is responsible for the CDW transition, just averaged over a large part of the Brillouin zone.

It is worth nothing that the constants A and B above are not adjustable parameters but are estimated from either neutron scattering or first principle calculations. A functional form very similar to Eq. 1 was proposed previously [49] based on an ad-hoc assumption that the composite excitation is a combination of a soft phonon with $q = q_{CDW}$ and another phonon from a non-soft branch with $q = -q_{CDW}$. However, this assumption lacks any physical justification so we do not consider it as a viable alternative. The dephasing time can be obtained by averaging the population numbers in a similar way,

$$\Gamma(T) = \Gamma_0 + \Gamma_{anh} \langle 1 + \frac{2}{\exp\left[\frac{\omega(T,q)}{T}\right] - 1} \rangle_q, \qquad (2)$$

where $\Gamma_0$ is the background contribution, $\Gamma_{anh}$ is the proportionality constant, and $n(\omega)$ is the thermal occupation number of the phonon mode. As shown in Fig. 6, the experimental data are well described by the proposed anharmonic decay model (Eq. 2).

Now we can turn to the analysis of the transient reflectivity dynamics. Transient optical response $\Delta A(\lambda)$ has been fitted using global fit approach in Glotaran [76] as follows:

$$\Delta A(\lambda) = \sum_i IRF \otimes DAS_i(\lambda) \times e^{-t/\tau_i} \qquad (3)$$

where the instrument response function (*IRF*) is being convoluted ($\otimes$) with global exponential decays with starting magnitudes of the decay associated spectra (*DAS$_i$*) and decay constants $\tau_i$ characterizing a particular relaxation channel. Typically, there are three distinct relaxation scales: $\tau_{e\text{-}e}$, the thermalization time of the electron subsystem due to electron-electron scattering, $\tau_{e\text{-}p}$, the time scale at which electron thermalize with the phonons *via* electron-phonon coupling, and $\tau_{diss}$, the time for the heat to dissipate away from the laser spot.

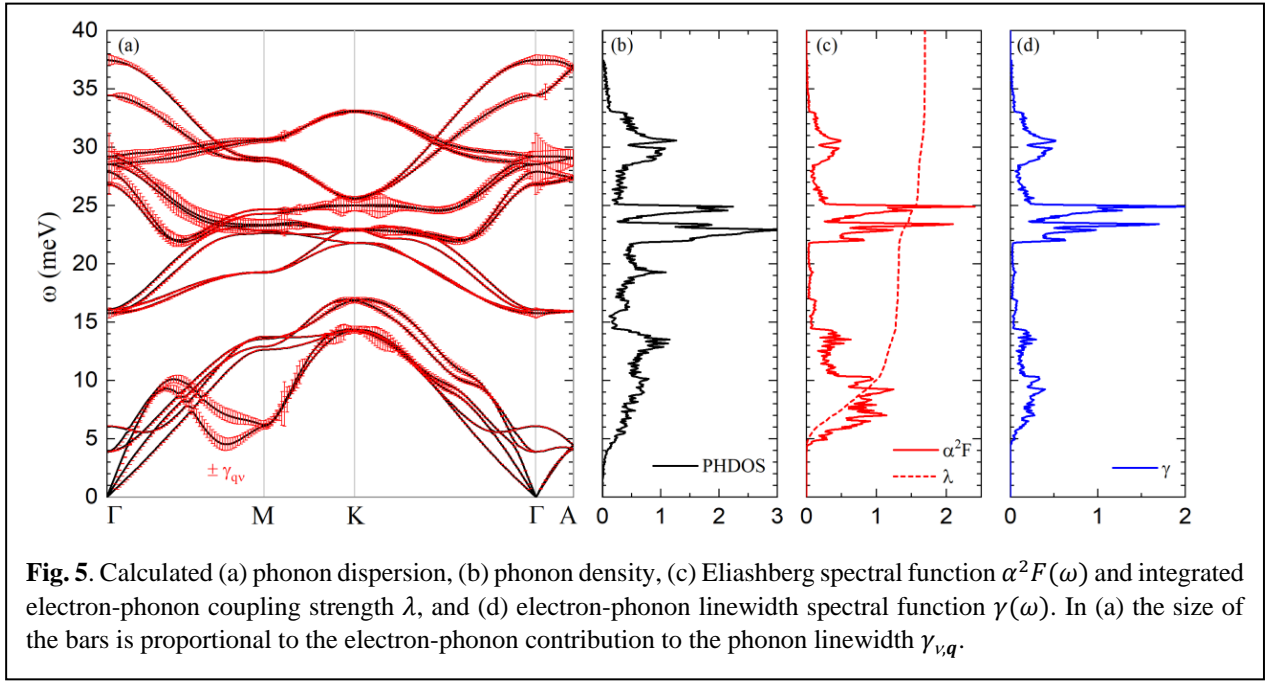

**Fig. 5.** Calculated (a) phonon dispersion, (b) phonon density, (c) Eliashberg spectral function $\alpha^2 F(\omega)$ and integrated electron-phonon coupling strength $\lambda$, and (d) electron-phonon linewidth spectral function $\gamma(\omega)$. In (a) the size of the bars is proportional to the electron-phonon contribution to the phonon linewidth $\gamma_{\nu,\mathbf{q}}$.

In our experiment, however, we can identify four different relaxation times, $\tau_i$, having temperature evolutions shown in Fig. 7(a-d). These timescales are on the order of ~100 fs, 1 ps, 100 ps and 1 ns. The fastest component, $\tau_1$, is extracted from the sharp negative peak that appears between 520 nm and 680 nm at the timescale of ~200 fs. Since the pump probe pulse has ~25 fs width, we cannot determine this relaxation time with a great accuracy although this peak has a significant negative amplitude. This is, obviously, the standard $\tau_{e\text{-}e}$ relaxation time.

The longest relaxation time $\tau_4$ is on the nanosecond scale and it is easily identifiable as $\tau_{diss}$. The temperature evolution shows a slowdown of the relaxation process at $T_{CDW}$ that is a signature of the second order CDW phase transition and is accompanied by an anomaly in elastic modulus and thermal conductivity at $T_{CDW}$. [77] At the highest fluence when CDW's electronic modulation is destroyed, the anomaly in $\tau_4$ at $T_{CDW}$ disappears.

Most interesting are the second and the third relaxation constants with 1 ps and 100 ps time scales. They have exponential temperature dependence, suggestive of electron-phonon and phonon-phonon scattering. In order to verify this assumption we have performed first-principle calculations of the electron-phonon coupling (see Supplementary Info). According to Allen's theory [78], relaxation of electrons with generation of a particular phonon $(\nu,\mathbf{q})$ of the branch $\nu$ with the wave vector $\mathbf{q}$ proceeds with the rate $1/\tau_{e-p} \propto \lambda_{\nu,\mathbf{q}} \omega_{\nu,\mathbf{q}}^2 = \gamma_{\nu,\mathbf{q}}$, where $\lambda_{\nu,\mathbf{q}}$ is the coupling constant with this particular mode and $\gamma_{\nu,\mathbf{q}}$ is the electron-phonon linewidth for the same mode. One may naively think that this process will be dominated by the soft mode with the

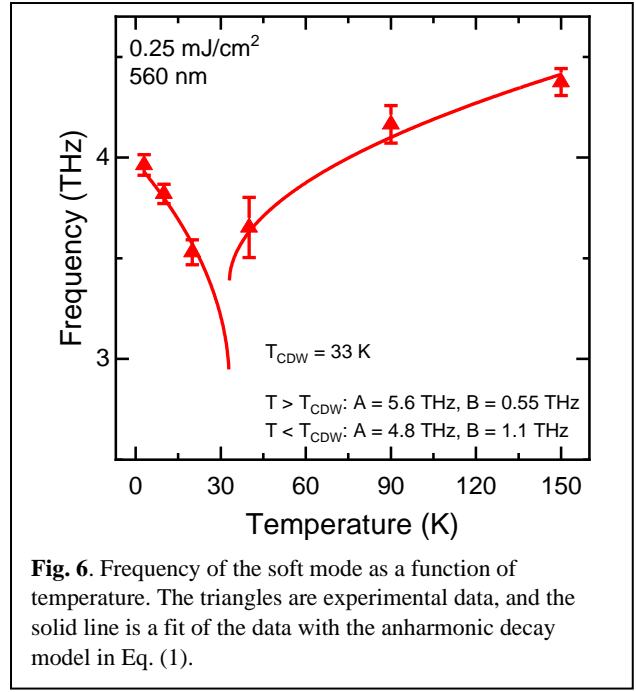

**Fig. 6.** Frequency of the soft mode as a function of temperature. The triangles are experimental data, and the solid line is a fit of the data with the anharmonic decay model in Eq. (1).

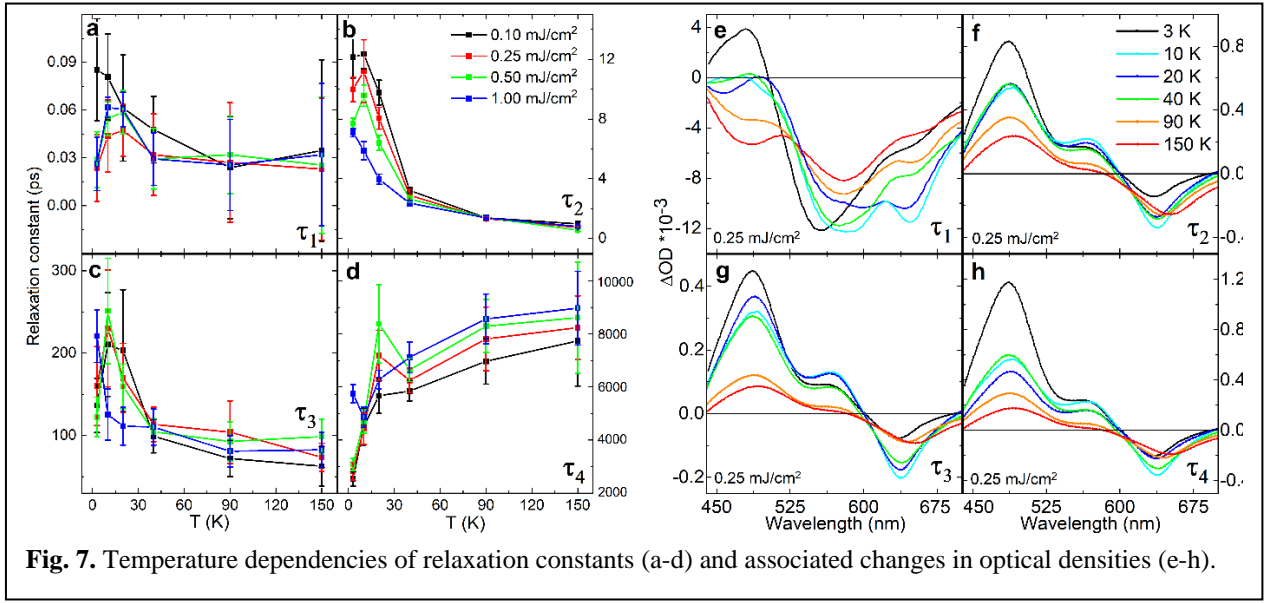
**Fig. 7.** Temperature dependencies of relaxation constants (a-d) and associated changes in optical densities (e-h).

large λ, but this is not the case. Rather, it is dominated by a group of phonons with frequencies of ~20 meV (4.8-5 THz).

We conclude that the *second* relaxation time, $\tau_2=\tau_{e\text{-}p}$, characterizes electron relaxation into these particular group of phonons, while the *third* one is due to, much slower, thermalization of these phonons with the rest of the phonon subsystem, *via* relatively slow, anharmonicity-driven phonon-phonon scattering. Thus, $\tau_3=\tau_{p\text{-}p}$.

Both $\tau_2$ and $\tau_3$ decay rapidly with temperature above $T_{CDW}$ which is a result of increase in phonon population number with temperature that leads to faster relaxation in both channels. As the temperature is lowered below $T_{CDW}$ we observe a decrease in relaxation times for all fluences, except for the highest one of 1 mJ/cm$^2$. This may be indirect evidence of electronic melting of the CDW state by a rapid flood of a large number of hot charge carriers, as observed in other CDW systems using pump-probe techniques [79,80].

Changes in optical density associated with each time constant are shown in Fig. 7(e-h). For $\tau_2$, $\tau_3$ and $\tau_4$ the spectral distribution has similar character – we observe features that correspond to transitions observed in static reflectivity at 485 nm (2.6 eV), 580 nm (2.15 eV), and 635 nm (1.95 eV) as described earlier in the text. At the transition near 485 nm, dramatic changes in optical density are seen when the sample enters the superconducting state. This is followed by a negligible change between 10 and 40 K within the CDW phase transition interval. Further detailed DFT calculation of the optical transitions are necessary to gain better understanding of the details in DAS(λ) amplitudes' behavior.

**Conclusions**

We have performed a broadband ultrafast reflectivity study of 2H-NbSe$_2$ in a wide temperature interval covering the normal, CDW and superconducting phase. Spectral features observed in the transient reflectivity experiment were associated with specific optical transitions obtained from band structure calculations. Displacive excitation of coherent phonons showed CDW-associated coherent oscillations of the soft phonon mode across the whole spectral range. Temperature evolution of this coherent phonon mode in the low-excitation linear regime shows softening of the mode down to $T_{CDW}$ with subsequent hardening below $T_{CDW}$. A global fit of the broadband probe data reveals four different relaxation times associated with characteristic electron-electron, electron-phonon and phonon-phonon relaxation times. From first principle calculations of electron-phonon coupling we associate the few ps electron-phonon relaxation time $\tau_2$ with a specific group of phonons with frequencies of ~20 meV. On the other hand, an anomalously long relaxation time of $\tau_3\sim100$ps is associated with anharmonicity-driven phonon-phonon scattering. All of the relaxation processes experience anomalies near the second order phase transition that are reflected in the temperature dependencies of the characteristic relaxation times and amplitudes of optical densities.

**Acknowledgements:**

A.A.A. was supported by the Center for Complex Materials from First Principles, an Energy Frontier Research Center funded by the U.S. Department of Energy, Office of Science, Basic Energy Sciences, under Grant No. DE-SC0012575. G.K acknowledges support from the NSF under Grant No. ECCS-1711015. E.R.M. acknowledges support from the National Science Foundation (Award No. OAC–1740263). This work used the Extreme Science and Engineering Discovery Environment (XSEDE) [81] which is supported by National Science Foundation grant number ACI-1548562. Specifically, this work used Comet at the San Diego Supercomputer Center through allocations TG–DMR180071. I.I.M. was supported by ONR through grant N00014-20-1-2345. Use of the Center for Nanoscale Materials, an Office of Science user facility, was supported by the U.S. Department of Energy, Office of Science, Office of Basic Energy Sciences, under Contract No. DE-AC02-06CH11357.